\documentclass[11pt]{article}
\usepackage{fullpage}
\usepackage{amsmath,amsfonts,amssymb}
\usepackage{nicefrac}

\usepackage{ifthen}
\usepackage{graphics,epsfig}
\usepackage{algpseudocode}
\usepackage{color}
\usepackage{algorithm}
\usepackage{authblk}

\newtheorem{theorem}{Theorem}

\newcommand{\qed}{\hfill $\Box$ \medbreak}
\newenvironment{proof}{\noindent {\bf Proof.}}{\qed}
\newtheorem{lemma}{Lemma}

\newcommand{\CONGEST}{{\sc congest}}
\newcommand{\ID}{\mbox{\rm ID}}
\newcommand{\accept}{\textsf{accept}}
\newcommand{\reject}{\textsf{reject}}
\newcommand{\child}{\mbox{\rm child}}
\newcommand{\act}{\mbox{\rm active}}
\newcommand{\col}{\mbox{\rm color}}
\newcommand{\cP}{\mathcal{P}}
\newcommand{\cT}{\mbox{\sc sos}}
\newcommand{\hF}{\hat{F}}
\newcommand{\hL}{\hat{L}}
\newcommand{\depth}{\operatorname{depth}}
\newcommand{\LOCAL}{{\sc local}}
\newcommand{\cO}{\mathcal{O}}

\begin{document}

\title{Distributed Subgraph Detection}
%\titlerunning{Distributed Subgraph Detection}

\author[1]{P. Fraigniaud\thanks{Additional support from ANR Project DESCARTES, and from INRIA Project GANG.}}
\author[2]{P. Montealegre\thanks{This work was partially supported by CONICYT via Basal in Applied Mathematics}}
\author[3]{D. Olivetti} 
\author[4]{I. Rapaport\thanks{This work was partially supported  by Fondecyt 1170021, N\'ucleo Milenio Informaci\'on y Coordinaci\'on en Redes ICM/FIC RC130003}}
\author[5]{I.~Todinca}

%\authorrunning{P. Fraigniaud, P. Montealegre, D. Olivetti, I. Rapaport, and I. Todinca}

\affil[1]{CNRS and Univ. Paris Diderot, France\\
  \texttt{pierre.fraigniaud@irif.fr}}

\affil[2]{Facultad de Ingenier\'{\i}a y Ciencias, Univ. Adolfo Ib\'a\~nez, Santiago, Chile\\
  \texttt{pedro.montealegre@uai.cl}}

\affil[3]{Gran Sasso Science Institute, L'Aquila, Italy\\
\texttt{dennis.olivetti@gssi.infn.it}}

\affil[4]{DIM-CMM  (UMI 2807 CNRS), Univ. de Chile, Santiago, Chile\\
  \texttt{rapaport@dim.uchile.cl}}

\affil[5]{Univ. Orl\'{e}ans, INSA Centre Val de Loire, LIFO EA 4022, Orl\'{e}ans, France\\
  \texttt{ioan.todinca@univ-orleans.fr}}
  
%\subjclass{F.2.2 Nonnumerical Algorithms and Problems; G.2.2 Graph Theory; F.1.2 Modes of Computation.}

%\keywords{Distributed computing, distributed decision, distributed property testing, \CONGEST\/ model.}

\iffalse
%Editor-only macros:: begin (do not touch as author)%%%%%%%%%%%%%%%%%%%%%%%%%%%%%%%%%%
\EventEditors{John Q. Open and Joan R. Acces}
\EventNoEds{2}
\EventLongTitle{42nd Conference on Very Important Topics (CVIT 2016)}
\EventShortTitle{CVIT 2016}
\EventAcronym{CVIT}
\EventYear{2016}
\EventDate{December 24--27, 2016}
\EventLocation{Little Whinging, United Kingdom}
\EventLogo{}
\SeriesVolume{42}
\ArticleNo{23}
% Editor-only macros::end %%%%%%%%%%%%%%%%%%%%%%%%%%%%%%%%%%%%%%%%%%%%%%%
\fi

\date{}
\maketitle

%%%%%%%%%%%%%%%%%%%%%%%%%%%%%%%%%%%%%%%%%%%%%%%
\begin{abstract}
In the standard \CONGEST\/ model for distributed network computing, it is known that ``global'' tasks such as minimum-weight spanning tree, diameter, and all-pairs shortest paths, consume rather large bandwidth, for their running-time is $\Omega(\mbox{poly}(n))$ rounds in $n$-node networks with constant diameter. Surprisingly, ``local'' tasks such as detecting the presence of a 4-cycle as a subgraph also requires $\widetilde{\Omega}(\sqrt{n})$  rounds (this bound holds even if one uses  randomized algorithms), and the best known upper bound for detecting  the presence of a 3-cycle is $\widetilde{O}(n^{\nicefrac23})$ rounds (randomized). The objective of this paper is to better understand the landscape of such  subgraph detection tasks. 
We show that, in contrast to \emph{cycles}, which are hard to detect in the \CONGEST\/ model, there exists a deterministic algorithm for detecting the presence of a subgraph isomorphic to $T$ running in a \emph{constant} number of rounds, for every \emph{tree}~$T$. Our algorithm  provides a distributed implementation of a combinatorial technique due to Erd\H{o}s et al.\! for sparsening the set of partial solutions kept by the nodes at each round. 

Our result has important consequences to \emph{distributed property-testing}, i.e., to randomized algorithms whose aim is to distinguish between graphs satisfying a property, and graphs far from satisfying that property. In particular, as a corollary of our result, we get that, for every graph pattern~$H$ composed of an edge and a tree connected in an arbitrary manner, there exists a (randomized) distributed testing algorithm for $H$-freeness, performing in  a constant number of rounds. Although the class of graph patterns $H$ formed by a tree and an edge connected arbitrarily may look artificial, all previous results of the literature concerning testing $H$-freeness for classical patterns $H$ such as cycles and cliques can be viewed as direct consequences of our result, while our algorithm enables testing more complex patterns. 
\end{abstract}
%%%%%%%%%%%%%%%%%%%%%%%%%%%%%%%%%%%%%%%%%%%%%%%

%%%%%%%%%%%%%%%%%%%%%%%%%%%%%%%%%%%%%%%%%%%%%%%
\section{Introduction}
%%%%%%%%%%%%%%%%%%%%%%%%%%%%%%%%%%%%%%%%%%%%%%%

%---------------------------------------------------------------------------
\subsection{Context and Objective}
%---------------------------------------------------------------------------

Given a fixed graph $H$ (e.g., a triangle, a clique on four nodes, etc.), a graph $G$ is $H$-free if it does not contain $H$ as a subgraph\footnote{Recall that $H$ is a subgraph of $G$ if $V(H)\subseteq V(G)$ and $E(H)\subseteq E(G)$}. Detecting copies of $H$ or deciding $H$-freeness has been investigated in many algorithmic frameworks, including classical sequential computing~\cite{ABK10}, parametrized complexity~\cite{MR808004}, streaming~\cite{buriol2006counting}, property-testing~\cite{alon1999efficient}, communication complexity~\cite{JS02}, quantum computing~\cite{AYZ94}, etc. In the context of distributed network computing, deciding $H$-freeness refers to the task in which the processing nodes of a network $G$ must collectively detect whether $H$ is a subgraph of~$G$, according to the following decision rule: 

\begin{itemize}
\item if $G$ is $H$-free then every node outputs \accept; 
\item otherwise, at least one node outputs \reject.
\end{itemize}

\noindent In other words, $G$ is $H$-free if and only if all nodes output \accept.

Recently, deciding $H$-freeness for various types of graph patterns $H$ has  received lots of attention (see, e.g., \cite{Censor-HillelFS16,Censor-HillelKKLPS2015,DLP2012,DruckerKO13,IzumiLG2017,FrOl17,FraigniaudRST16}) in the \CONGEST\/ model~\cite{Peleg2010}, and in variants of this model. (Recall that the \CONGEST\/ model is a popular model for analyzing the impact of limited link bandwidth on the ability to solve tasks efficiently in the context of distributed network computing). In particular, it has been observed that deciding $H$-freeness may require nodes to consume a lot of bandwidth, even for very simple graph patterns $H$. For instance, it has been shown in~\cite{DruckerKO13} that deciding $C_4$-freeness requires $\widetilde{\Omega}(\sqrt{n})$ rounds in $n$-node networks in the \CONGEST\/ model. Intuitively, the reason why so many rounds of computation are required to decide $C_4$-freeness is that the limited bandwidth capacity of the links prevents every node with high degree from sending the entire list of its neighbors through one link, unless consuming a lot of rounds. The lower bound $\widetilde{\Omega}(\sqrt{n})$ rounds for $C_4$-freeness can be extended to larger cycles $C_k$, $k\geq 4$, obtaining a lower bound of $\Omega(\mbox{poly}(n))$ rounds, where the exponent of the polynomial in~$n$ depends on~$k$~\cite{DruckerKO13}. Hence, not only ``global'' tasks such as minimum-weight spanning tree~\cite{SarmaHKKNPPW11,KuttenP98,OokawaI15}, diameter~\cite{AbboudCK16,FrischknechtHW12}, and all-pairs shortest paths~\cite{HolzerW12,LenzenP15,Nanongkai14} are bandwidth demanding, but also ``local'' tasks such as deciding $H$-freeness are bandwidth demanding, at least for some graph patterns $H$. 

In this paper, we focus on a generic set of $H$-freeness decision tasks which includes several instances deserving full interest on their own right. In particular, deciding $P_k$-freeness, where $P_k$  denotes the $k$-node path, is directly related to the \textsf{NP}-hard problem of computing the longest path in a graph. Also, detecting the presence of large complete binary trees, or of large binomial trees, is of interest for implementing classical techniques used in the design of efficient parallel algorithms (see, e.g., \cite{Leighton92}). Similarly, detecting large Polytrees in a Bayesian network might be used to check fast belief propagation~\cite{Pearl86}. Finally, as it will be shown in this paper, detecting the presence of various forms of trees can be used to tests the presence of graph patterns of interest in the framework of distributed property-testing~\cite{Censor-HillelFS16}. Hence, this paper addresses the following question: 

\begin{center}
\begin{minipage}{12cm}
For which tree $T$ is it possible to decide $T$-freeness efficiently in the  \CONGEST\/ model, that is, in a number of rounds independent from the size $n$ of the underlying network? 
\end{minipage}
\end{center}

At a first glance, deciding $T$-freeness for some given tree $T$ may look simpler than detecting cycles, or even just deciding $C_4$-freeness. Indeed, the absence of cycles enables to ignore the issue of checking that a path starts and ends at the \emph{same} node, which is bandwidth consuming for it requires maintaining all possible partial solutions corresponding to growing paths from all starting nodes. Indeed,  discarding even just a few starting nodes may result in missing the unique cycle including these nodes. However, even deciding $P_k$-freeness requires to overcome many obstacles. First, as mentioned before, finding a longest simple path in a graph is  \textsf{NP}-hard, which implies that it is unlikely that an algorithm deciding $P_k$-freeness exists in the \CONGEST\/ model, with running time polynomial in~$k$ at every node. Second, and more importantly, there  exists potentially up to $\Theta(n^k)$ paths of length~$k$ in a network, which makes impossible to maintain all of them in partial solutions, as the overall bandwidth of  $n$-node networks is at most $O(n^2\log n)$ in the \CONGEST\/ model. 
 
%---------------------------------------------------------------------------
\subsection{Our Results}
%---------------------------------------------------------------------------
 
We show that, in contrast to $C_k$-freeness, $P_k$-freeness can be decided in a constant number of rounds, for any $k\geq 1$. In fact, our main result is far more general, as it applies to \emph{any} tree. Stated informally, we prove the following: 

\medskip 

\noindent\textbf{Theorem A.} \textit{For every tree $T$, there exists a deterministic algorithm for \emph{deciding} $T$-freeness performing in a constant number of rounds under the \CONGEST\/ model. }
 
 \medskip 

For establishing Theorem~A, we present a distributed implementation of a pruning technique based on a combinatorial result due to Erd\H{o}s et al.~\cite{EHM64} that roughly states the following. Let $k>t>0$. For any set $V$ of $n$ elements, and any collection $F$ of subsets of~$V$, all with cardinality at most~$t$, let us define a \emph{witness} of $F$ as a collection $\widehat{F}\subseteq F$ of subsets of $V$ such that, for any $X\subseteq V$ with $|X|\leq k-t$, the following holds: 
 \[
\big ( \exists Y \in F : Y \cap X = \emptyset \big ) \; \Longrightarrow \;\big (\exists \widehat{Y} \in \widehat{F} :  \widehat{Y} \cap X = \emptyset \big). 
 \]
Of course, every $F$ is a witness of itself. However, Erd\H{o}s et al. have shown that, for every $k$, $t$, and $F$, there exists a \emph{compact} witness $\widehat{F}$ of $F$, that is, a witness whose cardinality depends on $k$ and $t$ only, and hence is independent of $n$.  To see why this result is important for detecting a tree~$T$ in a network $G$, consider $V$ as the set of nodes of $G$, $k$ as the number of nodes in $T$, and $F$ as a collection of subtrees $Y$ of size at most $t$, each isomorphic to some subtree of $T$. The existence of compact witnesses allows an algorithm to keep track of only a small subset $\widehat{F}$ of $F$. Indeed, if $F$ contains a partial solution $Y$ that can be extended into a global solution isomorphic to $T$ using a set of nodes $X$, then there is a \emph{representative} $\widehat{Y}\in\widehat{F}$ of the partial solution $Y\in F$ that can also be extended into a global solution isomorphic to $T$ using the same set $X$ of nodes. Therefore, there is no need to keep track of all partial solutions $Y\in F$, it is sufficient to keep track of just the partial solutions $\widehat{Y}\in\widehat{F}$. This pruning technique has been successfully used for designing fixed-parameter tractable (\textsf{FPT}) algorithms for the longest path problem~\cite{MR808004}, as well as, recently, for searching cycles in the context of distributed property-testing~\cite{FrOl17}.  Using this  technique for detecting the presence of a given tree however requires to push the recent results in~\cite{FrOl17} much further. First, the detection algorithm in~\cite{FrOl17} is anchored at a fixed node, i.e., the question addressed in~\cite{FrOl17} is whether there is a cycle $C_k$ passing through a given node. Instead, we address the detection problem in its full generality, and we do not restrict ourselves to detecting a copy of $T$ including some specific node. Second, detecting trees requires to handle partial solutions that are not only composed of sets of nodes, but that offer various shapes, depending on the structure of the tree~$T$, representing all possible combinations of subtrees of~$T$. 

Theorem~A, which establishes the existence of distributed algorithms for detecting the presence of trees, has important consequences on the ability to \emph{test} the presence of more complex graph patterns in the context of \emph{distributed property-testing}. Recall that, for $\epsilon\in(0,1)$, a graph $G$ is $\epsilon$-far from being $H$-free if removing less than a fraction $\epsilon$ of its edges cannot result in an $H$-free graph. A (randomized) distributed algorithm \emph{tests} $H$-freeness if it decides $H$-freeness  according to the following decision rule: 

\begin{itemize}
\item if $G$ is $H$-free then $\Pr[\mbox{every node outputs \accept}] \geq \nicefrac23$; 
\item if $G$ is $\epsilon$-far from being $H$-free then $\Pr[\mbox{at least one node outputs \reject}] \geq \nicefrac23$.  
\end{itemize}

That is, a testing algorithm separates graphs that are $H$-free from graphs that are far from being $H$-free. So far, the only non-trivial graph patterns $H$ for which distributed algorithms testing $H$-freeness are known are:

\begin{itemize}
\item the complete graphs $K_3$ and $K_4$ (see~\cite{Censor-HillelFS16,FraigniaudRST16}), and 
\item the cycles $C_k$, $k\geq 3$ (see~\cite{FrOl17}). 
\end{itemize}

Using our algorithm for detecting the presence of trees, we show the following (stated informally): 

\medskip 

\noindent\textbf{Theorem B.} \textit{For every graph pattern $H$ composed of an edge and a tree with arbitrary connections between them, there exists a (randomized) distributed  algorithm for \emph{testing} $H$-freeness performing in a constant number of rounds under the \CONGEST\/ model. }

\medskip

\begin{figure}[t]
\centering
 \includegraphics[width=0.6\textwidth]{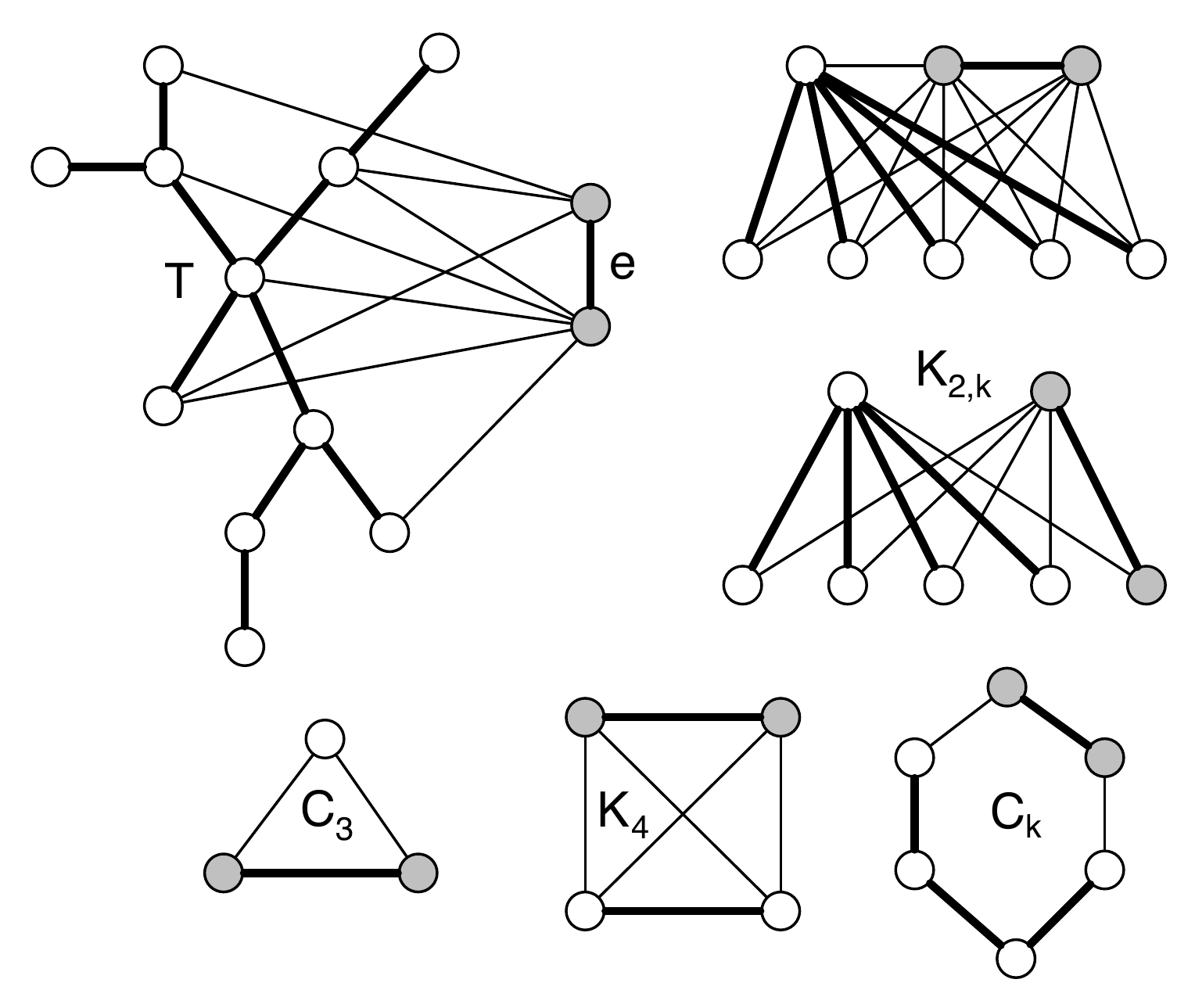}
\caption{All these graphs are composed of a tree $T$ and edge $e$ with arbitrary connections between them.}
\label{fig:TplusE}
\end{figure}

At a first glance, the family of graph patterns $H$ composed of an edge and a tree with arbitrary connections between them (like, e.g., the graph depicted on the top-left corner of Fig.~\ref{fig:TplusE}) may look quite specific and artificial. This is not the case. For instance, every cycle $C_k$ for $k\geq 3$ is a ``tree plus one edge''. This also holds for 4-node complete graph $K_4$. In fact, all known results about testing $H$-freeness for some graph $H$ in \cite{Censor-HillelFS16,FrOl17,FraigniaudRST16} are just direct consequence of Theorem~B. Moreover, Theorem~B enables to test the presence of other graph patterns, like the complete bipartite graph $K_{2,k}$ with $k+2$ nodes, for every $k\geq 1$, or the graph pattern depicted on the top-right corner of  Fig.~\ref{fig:TplusE}, in $O(1)$ rounds. It also enables to test the presence of connected 1-factors as a subgraph in $O(1)$ rounds. (Recall that a graph $H$ is a 1-factor if its edges can be directed so that every node has out-degree~1). 

In fact, our algorithm is 1-sided, that is, if $G$ is $H$-free, then all nodes output \accept\/ with probability~1. 

All our results are summarized on Table~\ref{table:summary}, together with some of the previous work in the literature. 

\begin{table}[h]
\begin{center}
\begin{tabular}{l|c|c|}
 & Distributed detection  & Distributed property testing  \\
\hline 
Cycles $C_k$ & $\Omega(\mbox{poly}(n))$ for $k\geq 4$ \cite{DruckerKO13}  & $O(1)$ \cite{Censor-HillelFS16,FrOl17,FraigniaudRST16} \\
\hline
Cliques $K_k$ &  $\widetilde{O}(n^{\nicefrac23})$ for $K_3=C_3$ \cite{IzumiLG2017} & $O(1)$ for $K_3$ and $K_4$ \cite{Censor-HillelFS16,FraigniaudRST16}  \\
 & open for $k\geq 4$  & open for $k\geq 5$   \\
\hline
Trees & $O(1)$ \textbf{[this paper]} & (same as left entry) \\
\hline
Trees-plus-one-edge & $\widetilde{\Omega}(\sqrt{n})$ for $C_4$ \cite{DruckerKO13} & $O(1)$  \textbf{[this paper]} \\
 & $O(\sqrt{n})$ for $C_4$ [Appendix] &  \\
\hline
Large pseudo-cliques & open & $O(1)$ \cite{BrakerskiP11} \\
\hline 
\end{tabular}
\end{center}
%\vspace{-3ex}
\caption{Number of rounds for deciding $H$-freeness in the \CONGEST\/ model}
\label{table:summary}
\end{table}
 
%---------------------------------------------------------------------------
\subsection{Previous Work}
%---------------------------------------------------------------------------

Subgraph detection has been the subject of a lot of investigations in the sequential computing setting. For the general problem of detecting whether a graph $H$ is a subgraph of $G$, where both $H$ and $G$ are part of the input, the best know bound is exponential \cite{Ullmann76}. Faster algorithms for special cases of graphs $H$ and $G$ are known. For example, if $H$ is a $k$-node tree, and $G$ is an $n$-node tree, then there is an $O(\frac{k}{\log k} n)$-time algorithm for deciding whether $H$ is a subgraph of $G$~\cite{ShamirT99}.  Subgraph detection becomes solvable in polynomial time if $H$ is fixed, and only $G$ is part of the input.  Moreover, for any fixed $H$, subgraph detection can be solved in linear time in planar graphs~\cite{Eppstein99}. In the  case of general graphs, but where $H=P_k$, the path of length~$k$, subgraph detection can be solved in time $O(n k!)$~\cite{MR808004}.
 
A relaxation of subgraph detection, called \textit{property testing} of subgraph freeness, aims at ``testing'' whether a graph $G$ given as input is $H$-free, by querying the nodes of the graphs at random. That is, the algorithm must distinguish between $H$-free graphs, and graphs that are $\epsilon$-far from being $H$-free. Several notion of $\epsilon$-farness have been introduced. In the \emph{dense} model (resp., \emph{sparse} model), a graph $G$ is $\epsilon$-far from satisfying a property if removing less than $\epsilon n^2$ edges (resp., $\epsilon m$ edges) of $G$ cannot result in a graph that satisfies the property. In the \emph{dense} model, the \emph{graph removal lemma} \cite{AlonFKS00,ConlonF2012,ErdosFR86} is exploited to test the presence of any fixed graph $H$ as subgraph (induced or not) in a constant number of queries. In the \emph{sparse} model, subgraph detection is harder. Even detecting triangles requires $\Omega(n^{\nicefrac{1}{3}})$ queries, and the best known upper bound is $O(n^{\nicefrac67})$ queries~\cite{AlonKKR08}. (The $\Omega(n^{\nicefrac{1}{3}})$ lower bound holds even for $2$-sided error algorithms, and for detecting any non bipartite subgraph). There exists a faster tester for cycle-detection in graphs of constant degree, as cycle-freeness can be tested with a constant number of queries by a  $2$-sided error algorithm~\cite{GoldreichR02}. However, testing cycle-freeness  using a $1$-sided error algorithms requires $\Omega(\sqrt{n})$ queries~\cite{CzumajGRSSS14}.

In the distributed setting, \cite{IzumiLG2017} very recently provided randomized algorithms for triangle detection, and triangle listing, in the \CONGEST\/ model, with round complexity $\widetilde{O}(n^{\nicefrac23})$ and $\widetilde{O}(n^{\nicefrac34})$, respectively, and establishes a lower bound $\widetilde{\Omega}(n^{\nicefrac13})$ on the round complexity of triangle listing.                                                       
Distributed property testing has been introduced in~\cite{BrakerskiP11}, where it is shown how to detect large pseudo-cliques in constant time. The topic has been recently reinvestigated and formalized in~\cite{Censor-HillelFS16}, for the \CONGEST\/ model. In this latter paper, it is shown that any sequential tester for the \emph{dense} model can be emulated in the \CONGEST\/ model, with just a quadratic slowdown (the number of rounds is the square of the number of queries). The same paper also provides distributed testers for triangle-freeness, cycle-freeness, and bipartiteness, in the sparse model, running in $O(1)$, $O(\log n)$, and $O(\mbox{polylog}\, n)$ rounds, respectively. In~\cite{FraigniaudRST16}, it is shown that, for every connected graph $H$ on four vertices, $H$-freeness can be tested in constant time. However, the same paper shows that the techniques used for testing $H$-freeness for 4-node graphs $H$ fail to test $C_k$-freeness or $K_k$-freeness in a constant number of rounds, whenever $k\geq 5$. It was recently shown in~\cite{FrOl17} that $C_k$-freeness can be tested in a constant number of rounds, for any $k\geq 3$.

Subgraph detection has also be investigated in the \emph{congested clique} model, a variant of the \CONGEST\/ model which separates the communication network (assumed to be a complete graph) from the input graph~$G$.  In~\cite{DLP2012}, it is shown that, for every $k$-node graph~$H$, deciding whether $H$ is a subgraph of an $n$-node input graph $G$ can be achieved in $\widetilde{O}(n^{1-2/k})$ rounds. Using an efficient implementation of parallel matrix multiplication algorithms in the congested clique, \cite{Censor-HillelKKLPS2015} improved the results in~\cite{DLP2012} for triangle detection (as well as for $C_4$-detection), via an algorithm running in $O(n^{0.158})$ rounds. 

Finally, \cite{DruckerKO13} studied subgraph detection in the \emph{broadcast} congested clique model, that is, the constrained variant of the congested clique model in which nodes are not allowed to send different messages to different neighbors in the clique. It is shown that, for every graph $H$, detecting whether the input graph $G$ contains $H$ as a subgraph can be done in $O(\frac{1}{n}\, \mbox{ex}(n,H))$ rounds, where $\mbox{ex}(n,H)$ denotes the Tur\'an number of $H$ and $n$. In term of lower bounds, it is proved in \cite{DruckerKO13} that detecting the clique $K_k$ requires $\widetilde{\Omega}(n)$ rounds for every $k\geq 4$,  detecting the cycle $C_k$ requires $\widetilde{\Omega}(\mbox{ex}(n,C_k)/n)$ rounds for every $k\geq 4$ (this result also holds for the  \CONGEST\/ model), and detecting the cycle $C_3$ requires $\widetilde{\Omega}(n/e^{O(\sqrt{\log n})})$ rounds. 

%---------------------------------------------------------------------------
\subsection{Structure of the paper}
%---------------------------------------------------------------------------

The \CONGEST\/ model is formally defined in the next section. Section~\ref{sec:detecttree} presents how to detect the presence of any tree in $O(1)$ rounds in this model. Section~\ref{sec:treeplusoneedge} presents the main corollary of this result, i.e., the ability to test the presence of any subgraph composed of a tree and an edge, with arbitrary connections between them,  in $O(1)$ rounds. Finally, Section~\ref{sec:conclusion} concludes the paper, by underlying some interesting research directions. 

(In addition, the Appendix presents a proof that the lower bound $\widetilde{\Omega}(\sqrt{n})$ rounds for deciding $C_4$-freeness in~\cite{DruckerKO13} is tight in the \CONGEST\/ model, up to polylog factors). 

%%%%%%%%%%%%%%%%%%%%%%%%%%%%%%%%%%%%%%%%%%%%%%%
\section{Model and notations}
%%%%%%%%%%%%%%%%%%%%%%%%%%%%%%%%%%%%%%%%%%%%%%%

In this paper, we use the classical \CONGEST\/ model for distributed network computing (see~\cite{Peleg2010}). We briefly recall the features of this model. The \CONGEST\/ model assumes a network modeled as a  connected simple (no self-loop, and no multiple edges) graph $G=(V,E)$. Each node $u\in V$ is provided with a $O(\log n)$-bit identity $\ID(u)$, and all identities are distinct. Nodes are honest parties, and links are reliable (i.e., the model is fault-free). All nodes starts at the same time, and computation proceeds in a sequence of synchronous rounds. At each round, every node sends messages to its neighbors in $G$, receives messages from these neighbors, and performs some individual computation. The messages sent at the same round can be different, although all our algorithms satisfy that, at every round, and for every node $u$, the messages sent by $u$ to its neighbors are identical. The are no limits on the computation power of the nodes. However, links are subject to a severe constraint: at every round, no more than $O(\log n)$ bits can traverse any given edge\footnote{Variants of the \CONGEST\/ model includes a parameter $B$, and no more than $B$ bits can be sent through an edge at any given round. In this paper, we stick to the classical variant in which $B=O(\log n)$.}. Hence, in particular, every node cannot send more than a constant number of  node IDs to each neighbor at every round. This makes the \CONGEST\/ model well suited to study network computing power limitation in presence of limited communication capacity, i.e., small link bandwidth. 

\subparagraph{Notation.} Given a network $G=(V,E)$, the set of neighbors of a node $u$ is denoted by~$N(u)$, and $\deg(u)=|N(u)|$ (recall that all considered graphs are simple). 

%%%%%%%%%%%%%%%%%%%%%%%%%%%%%%%%%%%%%%%%%%%%%%%
\section{Detecting the presence of trees}
\label{sec:detecttree}
%%%%%%%%%%%%%%%%%%%%%%%%%%%%%%%%%%%%%%%%%%%%%%%

In this section we establish our main result, i.e., Theorem~A, stated formally below as Theorem~\ref{th:tree}. As a warm up, we first show a simple and elegant randomized algorithm for deciding $T$-freeness, for every given tree~$T$, running in $O(1)$ rounds under the \CONGEST\/ model. Next, we show an algorithm that achieves the same, but deterministically.

%-------------------------------------------------------
\subsection{A simple randomized algorithm}
%-------------------------------------------------------

\begin{theorem}
For every tree $T$, there exists a 1-sided error randomized algorithm performing in $\cO(1)$ rounds in the \CONGEST\/ model, which correctly detects if the given input network contains $T$ as a subgraph, with probability at least $\nicefrac23$.
\end{theorem}

\begin{proof}
The algorithm performs in a sequence of phases. Algorithm~\ref{Tranddet} displays a phase of the algorithm. 

Let $k$ be the number of vertices of tree $T$, i.e., $k=|V(T)|$. Pick an arbitrary vertex of $T$, and root $T$ at that node. The root is labeled~$k$. Then, label the rest of the nodes of $T$ in decreasing order according to the order obtained from a BFS traversal starting from the root. For $i \in [1,k]$, let $T_i$ be the subtree of $T$ rooted at the node labeled~$i$. Let $\child(i)$ denote the labels of all the nodes adjacent to $i$ in $T_i$ (i.e., the labels of all the children of~$i$ in~$T$). We use the color coding technique introduced in~\cite{AYZ95} in the context of (classical) property testing. Each vertex  $u$ of $G$ picks a \emph{color} in $[1,k]$ uniformly at random. We say that $G$ is \emph{well colored} if at least one of the subgraphs $T'$ of $G$ that is isomorphic to $T$ satisfies that the colors of $T'$ correspond to the labels of the nodes in $T$. (Note that if $G$ is $T$-free then $G$ is well colored, no matter the coloring).  

In the verification algorithm, every vertex $u$ is either \emph{active} or \emph{inactive}, which is represented by a variable $\act(u) \in \{\mbox{true, false}\}$. Initially, every node $u$ is inactive (i.e., $\act(u)=$ false). Intuitively, a node $u$ becomes active if it has detected that the graph contains the tree $T_{c}$ as subgraph, rooted at $u$, where $c$ is the color of $u$. More precisely, once every node has picked a color in $[1,k]$ u.a.r., all nodes exchange their colors between neighbors. Then Algorithm~\ref{Tranddet} performs $k$ rounds. At the beginning of each round, every node $v$ communicates $\act(v)$ to all its neighbors. In round $c$, $1\leq c \leq k$, each node $u$ with color $c$ checks whether, for each color $c'$ of its children, some neighbor  $v$ is colored $c'$ and is active. If that is the case, it becomes active, otherwise it remains inactive. 

\begin{algorithm*}[h]
\caption{Randomized tree-detection, for a given tree $T$. Algorithm executed by node~$u$.}
\label{Tranddet}
\begin{algorithmic}[1] 
	\State{send $\ID(u)$ to all neighbors, and receive $\ID(v)$ from every neighbor $v$}
	\State{let $k=|V(T)|$, and pick $\col(u) \in [k]$ uniformly at random}
	\State{send $\col(u)$ to all neighbors, and receive $\col(v)$ from every neighbor $v$}
	\State{for every $c \in [1,k]$, let $N_c(u) =  \{v \in N(u) \mid \col(v) = c  \}$}
%	\If{$\child(\col(u))=\emptyset$} 
%		\State{$\act(u)\leftarrow $ true}
%		\Else 
%		\State{$\act(u) \leftarrow $ false}
%	\EndIf
	\State{$\act(u) \leftarrow $ false}
	\For{$c=1$ \textbf{to} $k$ }
		\State{send $\act(u)$ to all neighbors, and receive $\act(v)$ from every neighbor $v$}
		\State{compute $A(u) = \{v \in N(u) \mid  \act(v) = \mbox{true} \}$}
		\If{$\col(u)=c$ \textbf{and} $\big(\forall c' \in \child(c)$, $N_{c'}(u)\cap A(u) \neq \emptyset\big)$}
			\State{$\act(u) \leftarrow$ true} 
		\EndIf
	\EndFor
	\If{$\col(u) = k$ \textbf{and} $\act(u) =$ true}
	\State{output \reject}
	\Else
	\State{output \accept}
	\EndIf
	\end{algorithmic}
\end{algorithm*}

We claim that a well colored graph $G$ contains $T$ as a subgraph if and only if a vertex colored $k$ becomes active at round $k$. To establish that claim, note first that, if $c \in [1,k]$ is a leaf of $T$, then the tree $T_c$ is detected on round $c$, by every node colored $c$. Suppose now that, for every $c'<c$, the fact that a node $u$ colored $c'$ becomes active at round $c'$ means that $u$ has detected $T_{c'}$. Let $c_1, \dots, c_r$ be children of  $c$ in $T_c$. A node $u$ colored $c$ becomes active at round $c$ if and only, for every $i, 1\leq i \leq r$, it holds that $u$ has an active neighbor colored $c_i$. From the construction of the labels of $T$, and from the induction hypothesis, this implies that $u$ becomes active at round $c$ if and only if $u$ has detected $T_c$. We conclude that a node colored $k$ becomes active at round $k$ if and only if $T$ is detected in $G$, as $T=T_k$. 
 
Now, if $G$ contains $T$ as a subgraph, then the probability that $G$ is well colored is at least $(1/k)^k$.  Therefore, we run $\cO(k^k)$ independent iterations of  Algorithm~\ref{Tranddet}, which yields that, with probability at least $\nicefrac23$, $G$ is well colored for at least one iteration. 
\end{proof}

%-----------------------------------------------------------------------------------------
\subsection{Deterministic algorithm}
%-----------------------------------------------------------------------------------------

In this section, we establish our main result: 

\begin{theorem}\label{th:tree}
For every tree $T$, there exists an algorithm performing in $O(1)$ rounds in the \CONGEST\/ model for detecting whether the given input network contains $T$ as a subgraph. 
\end{theorem}

\begin{proof}
Let $k$ be the number of nodes in tree $T$. The nodes of $T$ are labeled arbitrarily by $k$ distinct integers in $[1,k]$. We arbitrarily choose a vertex $r\in[1,k]$ of $T$, and view $T$ as rooted in $r$. For any vertex $\ell \in V(T)$, let $T_\ell$ be the subtree of $T$ rooted in $\ell$. We say that $T_\ell$ is a \emph{shape} of $T$. Our algorithm deciding $T$-freeness proceeds in $\depth(T_r)+1$ rounds. At round~$t$, every node $u$ of $G$ constructs, for each shape $T_\ell$ of depth at most $t$, a set of subtrees of $G$ all rooted at $u$, denoted by $\cT_u(T_\ell)$, such that each subtree in $\cT_u(T_\ell)$ is isomorphic to the shape $T_\ell$. The isomorphism is considered in the sense of rooted trees, i.e., it maps $u$ to $\ell$. If we were in the \LOCAL\/ model\footnote{That is, the \CONGEST\/ model with no restriction on the size of the messages~\cite{Peleg2010}.}, we could afford to construct the set of all such subtrees of $G$. However, we cannot do that in the \CONGEST\/ model because there are too many such subtrees. Therefore, the algorithm acts in a way which guarantee that:  

\begin{enumerate}
\item the set  $\cT_u(T_\ell)$ is of constant size, for every node $u$ of $G$, and every node $\ell$ of $T$;
\item for every set $C \subseteq V$ of size at most $k-|V(T_\ell)|$, if there is some subtree $W$ of $G$ rooted at $u$ that is isomorphic to $T_\ell$, and that is not intersecting $C$, then $\cT_u(T_l)$ contains at least one such subtree $W'$ not intersecting $C$. (Note that $W'$ might be different from~$W$).
\end{enumerate}

The intuition for the second condition is the following. Assume that there exists some subtree $W$ of $G$ rooted at $u$, corresponding to some shape $T_\ell$, which can be extended into a subtree isomorphic to $T$ by adding the vertices of a set $C$. The algorithm may well not keep the subtree $W$ in  $\cT_u(T_l)$. However, we systematically keep at least one subtree $W'$ of $G$, also rooted at $u$ and isomorphic to $T_\ell$, that is also extendable to $T$ by adding the vertices of $C$. Therefore the sets $\cT_u(T_\ell)$, over all shapes $T_\ell$ of depth at most $t$, are sufficient to ensure that the algorithm can detect a copy of $T$ in $G$, if it exists. Our approach is described in Algorithm~\ref{al:Tdet}. (Observe that, in this algorithm, if we omit Lines~\ref{li:compact_begin} to~\ref{li:compact_end}, which prune the set $\cT_u(T_\ell)$, we obtain a trivial algorithm detecting $T$ in the \LOCAL\/ model). Implementing the pruning of the sets $\cT_u(T_\ell)$ for keeping them compact, we make use of the following combinatorial lemma, which  has been rediscovered several times, under various forms (see, e.g., \cite{FrOl17,MR808004}).

\begin{lemma}[Erd\H{o}s, Hajnal, Moon \cite{EHM64}]\label{pr:compactrepr}
Let $V$ be a set of size $n$, and consider two integer parameters $p$ and $q$. For any set $F \subseteq \cP(V)$ of subsets of size at most $p$ of $V$, there exists a \emph{compact $(p,q)$-representation} of $F$, i.e., a subset $\hF$ of $F$ satisfying:
\begin{enumerate}
\item For each set $C \subseteq V$ of size at most $q$, if there is a set $L \in F$ such that $L \cap C = \emptyset$, then there also exists $\hL \in \hF$ such that $\hL \cap C = \emptyset$;
\item The cardinality of $\hF$ is at most ${p+q \choose p}$, for any $n \geq p+ q$ .
\end{enumerate}
\end{lemma}

By Lemma~\ref{pr:compactrepr}, the sets  $\cT_u(T_\ell)$ can be reduced to constant size (i.e., independent of $n$), for every shape $T_\ell$ and every node $u$ of $G$. Moreover, the number of shapes is at most $k$, and, for each shape $T_\ell$, each element of $\cT_u(T_\ell)$ can be encoded on $k \log n$ bits. Therefore each vertex communicates only $O(\log n)$ bits per round along each of its incident edges. So, the algorithm does perform in $O(1)$ rounds in the \CONGEST\/ model\footnote{We may assume that, for compacting a set $\cT_u(T_\ell)$ in Lines~\ref{li:compact_begin}-\ref{li:compact_end},  every node $u$ applies Lemma~\ref{pr:compactrepr} by brute force (e.g., by testing all candidates $\hF$). In~\cite{MR808004}, an algorithmic version of Lemma~\ref{pr:compactrepr} is proposed, producing a set $\hF$ of size at most $\sum_{i=1}^q p^i$ in time $O((p+q)! \cdot n^3)$, i.e., in time $\mbox{poly}(n)$ for fixed $p$ and~$q$.}.

\begin{algorithm*}[h]
\caption{Tree-detection, for a given tree $T$. Algorithm executed by node~$u$.}
\label{al:Tdet}
	\begin{algorithmic}[1] 
	\For{each leaf $\ell$ of $T$}
		\State{let $\cT_u(T_\ell)$ be the unique tree with single vertex $u$}
		\State{exchange the sets $\cT$ with all neighbors}
	\EndFor
	\For{$t=1$ \textbf{to} $\depth(T)$}
		\For{each node $\ell$ of $T$ with $\depth(T_\ell) = t$}
		\label{line_for}
			\State{$\cT_u(T_\ell) \leftarrow \emptyset$}
			\State{let $j_1,\dots, j_s$ be the children of $\ell$ in $T$}
			\For{every $s$-uple $(v_1,\dots, v_s)$ of nodes in $N(u)$}
				\For{every  $(W_1,\dots,W_s)\in \cT_{v_1}(T_{j_1}) \times \dots \times  \cT_{v_s}(T_{j_s})$}
					\If{$\{u\}$ and $W_1,\dots,W_s$ are pairwise disjoint}
						\State{let $W$ be the tree with root $u$, and subtrees $W_1, \dots, W_s$}\label{li:glueing}
						\State{add $W$ to $\cT_u(T_\ell)$} \Comment{\textit{each $W_i$ is glued to $u$ by its root}}
					\EndIf
				\EndFor
			\EndFor
			\State{let $F = \{V(W) \mid W \in  \cT_u(T_l)\}$}\label{li:compact_begin}
			\Comment{\textit{collection of vertex sets for trees in $\cT_u(T_l)$)}}
			\State{construct  a $(|V(T_\ell)|,k-|V(T_\ell)|)$-compact representation $\hF \subseteq F$}
			\Comment{\textit{cf. Lemma~\ref{pr:compactrepr}}}
			\State{remove from $\cT_u(T_\ell)$ all trees $W$ with vertex set not in $\hF$}\label{li:compact_end}
			\State{exchange $\cT_u(T_\ell)$ with all neighbors}
		\EndFor
	\EndFor
	\If{$\cT_u(T_r) = \emptyset$} 	\Comment{\textit{$r$ denotes the root of $T$}}
		\State{accept}
	\Else
		\State{reject}
	\EndIf
	\end{algorithmic}
\end{algorithm*}

\subparagraph{Proof of correctness.} First, observe that if $\cT_u(T_\ell)$ contains a graph $W$, then $W$ is indeed a tree rooted at $u$, and isomorphic to $T_\ell$. This is indeed the case at round $t = 0$, and we can proceed by induction on $t$. Let $T_\ell$ be a shape of depth $\ell$. Each graph $W$ added  to $\cT_u(T_\ell)$ is obtained by gluing vertex-disjoint trees at the root $u$. These latter trees are isomorphic to the shapes $T_{j_1}, \dots, T_{j_s}$, where $j_1, \dots, j_s$ are the children of node $j$ in $T$. Therefore $W$ is isomorphic to $T_\ell$. In particular, if the algorithm rejects at some node $u$, it means that there exists a subtree of $G$ isomorphic to $T$. 

We now show that if $G$ contains a subgraph $W$ isomorphic to $T$, then the algorithm rejects in at least one node. For this purpose, we prove a stronger statement:

\begin{lemma}\label{le:avoid}
Let $u$ be a node of $G$, $T_\ell$ be a shape of $T$, and $C$ be a subset of vertices of $G$, with $|C| \leq k - |V(T_u)|$. Let us assume that there exists  a subgraph $W_u$ of $G$, satisfying the following two conditions:
(1)  $W_u$ is isomorphic to $T_\ell$, and the isomorphism maps $u$ on $\ell$, and 
(2)  $W_u$ does not contain any vertex of $C$.
Then $\cT_u(T_\ell)$ contains a tree $W'_u$ satisfying these two conditions.
\end{lemma}

We prove the lemma by induction on the depth of $T_\ell$. If $\depth(T_\ell) = 0$ then $\ell$ is a leaf of $T_\ell$, and $\cT_{u}(T_\ell)$ just contains the tree formed by the unique vertex $u$. Il particular, it satisfies the claim. 
Assume now that the claim is true for any node of $T$ whose subtree has depth at most $t-1$, and let $\ell$ be a node of depth $t$. Let $j_1,\dots, j_s$ be the children of $\ell$ in $T$. For every $i$, $1 \leq i \leq s$, let $v_i$ be the vertex of $W_u$ mapped on $j_i$. 
By induction hypothesis, $\cT_{v_1}(T_{j_1})$ contains some tree $W'_{v_1}$ isomorphic to $T_{j_1}$ and avoiding the nodes in $C \cup \{u\}$, as well as all the nodes of $W_{v_2}, \dots W_{v_s}$. Using the same arguments, we proceed by increasing values of $i=2,\dots,s$, and we choose a tree $W'_{v_i} \in \cT_{v_i}(T_{j_i})$ isomorphic to $T_{j_i}$ that avoids $C \cup \{u\}$, as well as all the nodes in $W'_{v_1}, \dots, W'_{v_{i-1}}$ and the nodes of $W_{v_{i+1}},\dots,W_{v_{s}}$. Now, observe that the tree $W''$ obtained from gluing $u$ to $W'_{v_1},\dots,W'_{v_s}$ has been added to $\cT_u(T_\ell)$ before compacting this set, by Line~\ref{li:glueing} of Algorithm~\ref{al:Tdet}. Since $W''$ does not intersect $C$, we get that, by compacting the set $\cT_u(T_\ell)$ using Lemma~\ref{pr:compactrepr}, the algorithm keeps a representative subtree $W'$ of $G$ that is isomorphic to $T_l$ and not intersecting $C$. This completes the proof of the lemma. \hfill $\diamond$ 

To complete the proof of Theorem~\ref{th:tree}, let us assume there exists a subtree $W$ of $G$ isomorphic to $T$, and let $u$ be the vertex that is mapped  to the root $r$ of $T$ by this isomorphism. By Lemma~\ref{le:avoid}, $\cT_u(T_r)\neq\emptyset$, and thus the algorithm rejects at node $u$.
\end{proof}

%%%%%%%%%%%%%%%%%%%%%%%%%%%%%%%%%%%%%% 
\section{Distributed Property Testing}
\label{sec:treeplusoneedge}
%%%%%%%%%%%%%%%%%%%%%%%%%%%%%%%%%%%%%% 

In this section, we show how to construct  a distributed tester for $H$-freeness in the sparse model, based on Algorithm \ref{al:Tdet}. This tester is able to test the presence of every graph pattern~$H$ composed of an edge $e$ and a tree $T$ connected in an arbitrary manner, by distinguishing graphs that include $H$ from graphs that are $\epsilon$-far\footnote{For $\epsilon\in(0,1)$, a graph $G$ is $\epsilon$-far from being $H$-free if removing less than a fraction $\epsilon$ of its edges cannot result in an $H$-free graph.} from being $H$-free. 

Specifically, we consider the set $\cal H$ of all graph patterns $H$ with node-set $V(H) = \{x, y, z_1, \dots, z_k \}$ for $k\geq 1$, and edge-set $E(H) =  \{f\}   \cup E(T) \cup \cal E$, where $f=\{x,y\}$, $T$~is a tree with node set $\{z_1, \dots, z_k\}$, and $\cal E$ is some set of edges with one end-point equal to $x$ or $y$, and the other end-point $z_i$ for $i\in\{1,\dots,k\}$. Hence, a graph $H\in\cal H$ can be described by a triple $(f,T,\cal E)$ where $\cal E$ is a set of edges connecting a node in $T$ with a node in~$f$. 

We now establish our second main result, i.e., Theorem~B, stated formally below as follows: 

\begin{theorem}
For every graph pattern $H\in \cal H$, i.e., composed of an edge and a tree connected in an arbitrary manner, there exists a randomized $1$-sided error distributed property testing algorithm for $H$-freeness performing in $O(\nicefrac{1}{\epsilon})$ rounds in the \CONGEST\/ model. 
\end{theorem}

\begin{proof}
Let $H=(f,T,\cal E)$, with $f=\{x,y\}$. Let us assume that there are $\nu$ copies of $H$ in $G$, and let us call these copies $H_1 = (f_1,T_1,{\cal E}_1), \dots, H_\nu = (f_\nu,T_\nu,{\cal E}_\nu))$. Let $\mathbf{E} = \{ f_1, \ldots, f_\nu\}$. Our tester algorithm for $H$-freeness is composed by the following two phases:

\begin{enumerate}
	\item determine a candidate edge $e$ susceptible to belong to $\mathbf{E}$;
	\item checking the existence of a tree $T$ connected to $e$ in the desired way.
\end{enumerate}

In order to find the candidate edge, we exploit the following lemma:

\begin{lemma}[\cite{FraigniaudRST16}] \label{disjointcopies}
Let $H$ be any graph. Let $G$ be an $m$-edge graph that is $\epsilon$-far from being $H$-free. Then $G$ contains at least $\epsilon m / |E(H)|$ edge-disjoint copies of $H$.
\end{lemma}

Hence, if the actual $m$-edge graph $G$ is $\epsilon$-far from being $H$-free, we have $|\mathbf{E}|\geq \epsilon m /  |E(H)|$. Thus, by randomly choosing an edge $e$ and applying Lemma \ref{disjointcopies}, $e\in\mathbf{E}$ with probability at least $\epsilon /  |E(H)|$. 

As shown in \cite{FrOl17}, the first phase can be computed in the following way. First, every edge is assigned to the endpoint having the smallest identifier. Then, every node picks a random integer $r(e)\in[1,m^2]$ for each edge $e$ assigned to it. The candidate edge of Phase~1 is the edge $e_{min}$ with minimum rank, and indeed
$
\Pr[e_{min}\in\mathbf{E}]\geq \epsilon /  |E(H)|.
$

It might be the case that $e_{min}$ is not unique though. However: 
$
\Pr[e_{min} \; \mbox{is unique}] \geq \nicefrac{1}{e^2}
$
where $e$ denotes here the basis of the natural logarithm. Also, every node picks, for every edge $e=\{v_1,v_2\}$ assigned to it, a random bit~$b$. Assume, w.l.o.g., that $\ID(v_1)<\ID(v_2)$. If $b=0$, then the algorithm will start Phase~2 for testing the presence of $H$ with $(x,y)=(v_1,v_2)$, and if $b=1$, then the algorithm will start Phase~2 for testing the presence of $H$ with $(x,y)=(v_2,v_1)$. We have
$
\Pr[e_{min} \, \mbox{is considered in the right order}] \geq \nicefrac{1}{2}. 
$
It follows that the probability $e_{min}$ is unique, considered in the right order, and part of $\mathbf{E}$ is at least $\frac{\epsilon}{2|E(H)| e^2}$. 

Using a deterministic search based on Algorithm~\ref{al:Tdet}, $H$ will be found with probability at least $\frac{\epsilon}{2|E(H)| e^2}$. To boost the probability of detecting $H$ in a graph that is $\epsilon$-far from being $H$-free, we repeat the search $2 e^2 |E(H)| \ln 3 / \epsilon$ times. In this way, the probability that $H$ is detected in at least one search is at least $2/3$ as desired.

During the second phase, the ideal scenario would be that all the nodes of $G$ search for $H=(f,T,\cal E)$ by considering only the edge $e_{min}$ as candidate for $f$, to avoid congestion. Obviously, making all nodes aware of $e_{min}$ would require diameter time. However, there is no needs to do so. Indeed, the tree-detection algorithm used in the proof of Theorem~\ref{th:tree} runs in $depth(T)$ rounds. Hence, since only the nodes at distance at most $depth(T)+1$ from the endpoints of $e_{min}$ are able to detect $T$, it is enough to broadcast $e_{min}$ at distance up to $2 \, (depth(T)+1)$ rounds. This guarantees that all nodes participating to the execution of the algorithm for $e_{min}$ will see the same messages, and will perform the same operations that they would perform by executing the algorithm for $e_{min}$ on the full graph. So, every node broadcasts its candidate edges with the minimum rank, at distance  $2 \, (depth(T)+1)$. Two contending broadcasts, for two candidate edges $e$ and $e'$ for $f$, resolve contention by discarding the broadcast corresponding to the edge $e$ or $e'$ with largest rank. (If $e$ and $e'$ have the same rank, then both broadcast are discarded). After this is done, every node is assigned to one specific candidate edge, and starts searching for $T$. Similarly to the broadcast phase, two contending searches, for two candidate edges $e$ and $e'$, resolve contention by aborting the search corresponding to the edge $e$ or $e'$ with largest rank. From now on, one can assume that a single search in running, for the candidate edge $e_{min}$. 

It remains to show how to adapt Algorithm~\ref{al:Tdet} for checking the presence of a tree $T$ connected to a \emph{fixed} edge $e = \{x,y\}\in E(G)$ as specified in $\cal E$. Let us consider Instruction~\ref{line_for} of Algorithm \ref{al:Tdet}, that is: ``\textbf{for} each node $\ell$ of $T$ with $\depth(T_\ell) = t$ \textbf{do}''.
At each step of this for-loop, node $u$ tries to construct a tree $W$ that is isomorphic to the subtree of $T$ rooted at $\ell$. In order for $u$ to add $W$ to $\cT_u(T_\ell)$, we add the condition that:
\begin{itemize}
\item if $\{\ell,x\} \in E(H)$ then $\{u,x\} \in E(G)$, and
\item if $\{\ell,y\} \in E(H)$ then $\{u,y\} \in E(G)$. 
\end{itemize}
Note that this condition can be checked by every node $u$. If this condition is not satisfied, then $u$ sets $\cT_u(T_\ell)=\emptyset$.

This modification enables to test $H$-freeness. Indeed, if the actual graph $G$ is $H$-free, then, since at each step of the modified algorithm, the set $\cT_u(T_\ell)$ is a subset of the set $\cT_u(T_\ell)$ generated by the original algorithm, the acceptance of the modified algorithm is guaranteed from the correctness of the original algorithm. 

Conversely, let us show that, in a graph $G$ that is $\epsilon$ far of being $H$-free, the algorithm rejects $G$ as desired. In the  first phase of the algorithm, it holds that $e_{min}\in \mathbf{E}$ happens in at least one search whenever $G$ is $\epsilon$-far from being $H$-free, with probability at least $\nicefrac{2}{3}$. Following the same reasoning of the proof of Lemma \ref{le:avoid}, since the images of the isomorphism satisfy the condition of being linked to nodes $\{x,y\}$ in the desired way, the node of $G$ that is mapped to the root of $T$ correctly detects $T$, and rejects, as desired.
\end{proof}

%%%%%%%%%%%%%%%%%%%%%%%%%%%%%%%%%%%%%%%%%%%%%%%
\section{Conclusion}
\label{sec:conclusion}
%%%%%%%%%%%%%%%%%%%%%%%%%%%%%%%%%%%%%%%%%%%%%%%
 
In this paper, we have proposed a generic construction for designing deterministic distributed algorithms detecting the presence of any given tree $T$ as a subgraph of the input network, performing in a constant number of rounds in the \CONGEST\/ model. Therefore, there is a clear dichotomy between cycles and trees, as far as efficiently solving $H$-freeness is concerned: while every cycle of at least four nodes requires at least a polynomial number of rounds to be detected, every tree can be detected in a constant number of rounds. It is not clear whether one can provide a simple characterization of the graph patterns $H$ for which $H$-freeness can be decided in $O(1)$ rounds in the \CONGEST\/ model. Indeed, the lower bound $\widetilde{\Omega}(\sqrt{n})$ for $C_4$-freeness  can be extended to some graph patterns containing $C_4$ as induced subgraphs. However, the proof does not seem to be easily extendable to all such graph patterns as, in particular, the patterns containing many overlapping $C_4$ like, e.g., the 3-dimensional hypercube $Q_3$, since this case seems to require non-trivial extensions of the proof techniques in~\cite{DruckerKO13}. An intriguing question is to determine the round-complexity of deciding $K_k$-freeness in  the \CONGEST\/ model for $k\geq 3$, and in particular to determine the exact round-complexity of deciding $C_3$-freeness.

Our construction also provides randomized algorithms for testing $H$-freeness (i.e., for distinguishing $H$-free graphs from graphs that are far from being $H$-free), for every graph pattern $H$ that can be decomposed into an edge and a tree, with arbitrary connections between them, also running in $O(1)$ rounds  in  the \CONGEST\/ model. This generalizes the results in~\cite{Censor-HillelFS16,FrOl17,FraigniaudRST16}, where  algorithms for testing $K_3$, $K_4$, and $C_k$-freeness for every $k\geq 3$ were provided. Interestingly, $K_5$ is the smallest graph  pattern $H$ for which it is not known whether testing $H$-freeness can be done in $O(1)$ rounds, and this is also the smallest graph pattern that cannot be decomposed into a tree plus an edge.  We do not know whether this is just coincidental or not.  

\bibliographystyle{plain}
\bibliography{Biblio}

%%%%%%%%%%%%%%%%%%%%%%%%%%%%%%%%%%%%%%%%%%%%%%%
\appendix
\centerline{\bf \Large APPENDIX}
\section{On detecting $C_4$ in the \CONGEST\/ model}
%%%%%%%%%%%%%%%%%%%%%%%%%%%%%%%%%%%%%%%%%%%%%%%

Detecting the presence of $C_3$ as subgraph can be done in $\widetilde{O}(n^{\nicefrac23})$ rounds by a randomized algorithm~\cite{IzumiLG2017}, but the exact round complexity of detecting $C_3$ is not known (no non-trivial lower bound). On the other hand, a non-trivial lower bound is known about detecting the presence of $C_4$ as subgraph:

\begin{theorem}[\cite{DruckerKO13}]\label{theo:lowerbound}
There are no algorithms for $C_4$-freeness in $n$-node networks performing in less than $\Omega(\sqrt{n}/\log n)$ rounds  in the \CONGEST\/ model. 
\end{theorem}

%\begin{proof}
%Recall that set-disjointness $\mbox{DISJ}(N)$ is a 2-party communication complexity problem where Alice is given a subset $A$ of some $N$-element ground set $U$,  Bob is given a subset $B$ of the same set $U$, and they must decide whether $A\cap B=\emptyset$ or not. The seminal result in~\cite{Yao79} states that $\mathsf{CC}(\mbox{DISJ}(N))=\Omega(N)$, that is, any protocol enabling Alice and Bob to decide set-disjointness requires the exchange of $\Omega(N)$ bits between the two parties. The reduction from set-disjointness to $C_4$-freeness proceeds as follows. Let $H$ be a densest $n$-node $C_4$-free graph. It is known~\cite{ERS66} that there exists $H$ with $N=\Omega(n^{\nicefrac32})$ edges. Let $G$ be the graph composed of two copies of $H$ with a perfect matching between sibling nodes. The sets $A$ and $B$ given to Alice and Bob in $\mbox{DISJ}(N)$ can be viewed as two subsets of the ground set $U=E(H)$. The graph $G_{A,B}$ is defined as the graph $G$ where all edges in $E(H)\setminus A$ are removed from one copy of $H$, and all edges in $E(H)\setminus B$ are removed from the other copy of $H$. We have that $G_{A,B}$ is $C_4$-free if and only if $A\cap B=\emptyset$. The total bandwidth available between the two copies of $H$ is $O(n\log n)$ bits. Therefore, $C_4$-freeness cannot be decided in less than $\Omega(\frac{n^{\nicefrac32}}{n\log n})=\Omega(\sqrt{n}/\log n)$. 
%\end{proof}

Interestingly, the lower bound of Theorem~\ref{theo:lowerbound} is tight up to log factors, using a very simple algorithm. 

\begin{theorem}
There exists an algorithm performing in $O(\sqrt{n})$ rounds in the \CONGEST\/ model for solving $C_4$-freeness in $n$-node networks. 
\end{theorem}

\begin{proof}
A $O(\sqrt{n})$-rounds algorithm for $C_4$-detection is displayed in Algorithm~\ref{c4det}. We prove its correctness. It was observed in~\cite{Censor-HillelKKLPS2015} that if a node $u$ satisfies $\sum_{v\in N(u)}\deg(v)\geq 2n+1$, then $u$ belongs to a $C_4$. Hence, Instruction~\ref{c4det:degree} correctly detects such a 4-cycle. Therefore, we can now assume, w.l.o.g., that every node $u$ satisfies  $\sum_{v\in N(u)}\deg(v)\leq 2n$. It follows from this assumption that no nodes can have more than $\sqrt{2n}$ heavy neighbors, where a node is \emph{heavy} if it has degree more than  $\sqrt{2n}$, and it is \emph{light} otherwise. As a consequence, every heavy neighbor $w$ of every node $v$ belongs to $S(v)$. So, if there exists a 4-cycle $(u,v_1,w,v_2)$ where $w$ is an heavy node, then the two neighbors $v_1$ and $v_2$ of $w$ on this cycle will send $\ID(w)$ to $u$, leading $u$ to correctly reject in Instruction~\ref{c4det:reject}. Finally, if there exists a 4-cycle composed of solely light nodes, then all nodes of that cycle will correctly reject in Instruction~\ref{c4det:reject} because each of them sends the IDs of all its neighbors to each of its neighbors. 
\end{proof}

\begin{algorithm*}[]
\caption{$C_4$-detection executed by node $u$.}
\label{c4det}
\begin{algorithmic}[1] 
	\State{send $\ID(u)$ to all neighbors, and receive $\ID(v)$ from every neighbor $v$}
	\State{send $\deg(u)$ to all neighbors, and receive $\deg(v)$ from every neighbor $v$}
	\State{$S(u)\leftarrow \{$IDs of the $\min\{\sqrt{2n},\deg(u)\}$ neighbors with largest degrees$\}$}
	\State{send $S(u)$ to all neighbors, and receive $S(v)$ from every neighbor $v$}
	\If{$\sum_{v\in N(u)}\deg(v)\geq 2n+1$} 
		\State{output \reject} \label{c4det:degree}
	\Else
		\If{$\exists v_1,v_2\in N(u), \exists w \in S(v_1)\cap S(v_2): w \neq u \; \mbox{and} \;  v_1\neq v_2$} 
			\State{output \reject} \label{c4det:reject}
		\Else
			\State{output \accept}
		\EndIf		
	\EndIf		
	\end{algorithmic}
\end{algorithm*}

%%%%%%%%%%%%%%%%%%%%%%%%%%%%%%%%%%%%%%%%%%%%%%%
\end{document}